\documentstyle[12pt]{article}
\def\arr{\rightarrow}
\begin{document}
\rightline{IASSNS-HEP-95/84}
\begin{center} {\Large \bf Some Speculations on the Ultimate Planck Energy
Accelerators}
\end{center}
\bigskip \bigskip
\begin{center}
A. Casher\\
Tel Aviv University

S. Nussinov\\
Tel Aviv University \\
and
Institute for Advanced Study\\
Princeton, NJ 08540

\end{center}
\section{Abstract}

 The inability to achieve in the present universe, via electromagnetic
 or gravitational acceleration, Planck energies for elementary
particles is suggested on the
basis of several, some relatively sophisticated,failed attempts.
 This failure is essential for schemes
were the superplanckian regime for the energies of elementary particles
is ``Unphysical''.The basic observation is that this failure to achieve
superplanckian energies naturally
occurs in our universe of finite age and horizon. It does tie up in
a  mysterious fashion these cosmological quantities
and elementary physics parameters such as the masses of the lightest
charged fermions.

\section{Introduction}
The Planck mass, $m_P = (\hbar c/G_{Newton})^{1/2} \approx 10^{19}\; GeV$ and
corresponding length $l_P = \hbar/m_Pc \approx 10^{-33} cm$  or time $t_P =
l_P/c$, are of fundamental importance, marking the onset of strong
non-renormalizable quantum gravity effects.
 Planck mass objects could be the end products of black holes emitting
Hawking radiation\cite{Haw}, obtained when the radius, mass and temperature of
the black hole become Planckian.  At this point the semiclassical argument for
the radiation -- based in particular on the existence of a well defined
horizon breaks down due to quantum metric--fluctuations.Also the total
mass of the black hole, $m_P$, disallows emission
of further quanta with $E \approx T \approx m_P$.  Stable Planck mass
objects have been offered \cite{Aha} as a possible solution to the apparent
unitarity (or information) crisis encountered if the black hole completely
evaporates.It has been conjectured \cite{Cas} that a reciprocal
mechanism exists, which
protects mini black holes from crossing the $M_{BH} = m_P$ line downwards,
and prevents elementary particles (field quanta) from crossing the
$m_{particle} = m_P$ line upwards. A possible realization for this mechanism
calls for accumulation of states at $m = M = m_P$.

A new, radical, approach of P. Mazur\cite {Maz} attempts to side-step the
problems of non-renormalize\-ability due to elementary super-planckian
excitations \underline{and} of collapsing mini-black holes, in a kinematic
rather than a dynamic way.  The Planck length is introduced into a new set of
commutators $[x,y] \sim [z,t] \sim il{_P}{^2}$.  The corresponding uncertainty
relations {\it e.g.} $\Delta r \Delta t \geq {l_P}{^2}/2$ excludes the
problematic domain of one gravitating quantum degree of freedom when the latter
is inside its own Schwartzschild radius -- [just like $[x,p] = i\hbar$ and
Heisenberg's uncertainty avoids the singular spiralling to the origin of a
radiating electron in the classical Coulomb problem].  The (area)
discretization implied by the new commutators suggests a finite difference
analogue of Schr\"{o}dinger's equation for the self gravitating
quantum bubble \cite{Ber} that indeed implements the above exclusion.
  The finite difference  coordinate
space equation implies periodicity in momentum space with a period $\sim m_P$.
 The Planck mass $m_P$ is the limiting value of momentum or energy
of any (elementary) particle.\footnote{Clearly, massive extended macroscopic
objects are irrelevant, as the quantum gravity divergences come from
concentrating $m_P$ energy densities within $l_P$ size.  The proton is
composite, and so could be the electron (and quark).  We are assuming, however,
finite compositeness: 3 quarks, finite \# of'' preons'', etc., which suffices
for
our purpose.}  This is clearly stronger than the assumption  that no
two oppositely moving transplanckian
(elementary) particles can collide in an S wave to form a
mini-black hole of mass $M_{BH} \geq m_P$.$^1$
\section{The Implications of the Maximal P, E Postulate}
The suggestion that one cannot boost an electron or proton to super-planckian
energies flies in the face of Lorentz invariance , one of the best
tested principles in nature.  It has been
argued\cite{Maz} that due to a discrete coordinate structure underlying space
time at short $\approx l_P$ scales, there could be -- as in ordinary crystals
--
``Umklapp" processes which ``absorb" a reciprocal lattice momentum, keeping
$E, P \leq m_P$.  Nonetheless, the following gedanken experiment points at some
difficulty.
   Suppose that a proton is accelerated inside some microscopic device $D_1$
(see
fig 1) to $\gamma = 11$.  The whole $D_1$ device in turn sits within a
larger setup $D_2$, which boosts $D_1$ to $\gamma = 11$.  $D_2$ in turn sits
inside $D_3$... etc., etc.  The device $D_{18}$ is then also boosted by
$D_{19}$ to $\gamma = 11$, thus finally achieving transplanckian energies $E
= \gamma^{19}\, GeV \approx\, 10^{20}\, GeV$ for our proton.  To avoid
this we need to assume that the last device $D_{19}$ ``knows" that eighteen
layers down, inside the innermost microscopic $D_1$, \underline{one proton} is
about to break the ``Planck Barrier".{This possibility is even more unlikely
than the Grimm brothers' fairy tale about the sensitive princess who could not
fall asleep due to one pea under seventeen mattresses.} It would be truly
paradoxical if space-time on large scales was completely flat
and uniform.  However, precisely due to gravity $(G_N > 0)$ and finite
propagation velocity $(c < \infty)$, the universe is curved and only a finite
horizon has opened up since the big bang.  To maintain the ``Planck Barrier",
an
ultimate micro-macro connection may be at work.  It should force the largest
device $D_{19}$ in our gedanken experiment to be larger than the
 observed universe, and cause any alternative less massive and/or more
compact design to break due to finite strength of materials, (an issue
involving $\bar{h} > 0$) or collapse into a black hole.

\section{Attempts to Build (Gedanken) Super-planckian Accelerators}

  By
considering a few gedanken accelerating devices, we would like to suggest that
 $E > m_P$ may be inaccessible in our
universe.

\noindent {\large \bf (a) Electromagnetic acceleration methods}\\
{\bf (1) Laser Beam Acceleration}:  An intense laser beam can accelerate
charged particle to high energies by repeated Compton scattering.  As the
particle approaches the
putative super-planckian regime, it becomes extremely relativistic.  In
the particle's rest frame the photons will be strongly redshifted (by a
$\gamma^{-1}$ factor) and $\sigma = \sigma_{Thompson} \approx \pi \alpha^2/m^2$
is appropriate.  If we have a flux of energy $\Phi_E$ the particle gains
energy at a rate:
\begin{equation}
\frac{dW}{dt} = \Phi_E \sigma (1 - \beta ) = \Phi_E \frac{\pi \alpha^2}{m^2}
\frac{m^2}{2m{_P}{^2}} \approx \Phi_E \frac{\alpha^2}{m{_P}{^2}}
\end{equation}

The key factor entering the rate is the relative velocity $(1 - \beta ) \approx
\frac{1}{2\gamma^2}$.  Thus even if we allow Hubble time acceleration $t_H
\approx 3.10{^{17}}{_{Sec}}$, we need, regardless of the mass $m$ of the
accelerated particle, a minimal energy flux  $\Phi_E =
\frac{m_P}{\alpha^2l{_P}{^2}t_H} \approx
3.10^{68} \frac{ergs}{cm^2 sec}$.  The corresponding E.M. energy density is $U
\approx 10^{58}\, \frac{ergs}{cm^3} \approx 3. 10^{41}\, (\frac{ergs}{cm})^2$.
Using $U \approx E^2/8\pi$ we find that the E field in the beam should exceed
$\approx 10^{34}$ Volts/cm.  Such fields are completely untenable.
The vacuum itself ``sparks" and produces $e^+e^-$ pairs at a rate\cite{Sch}:
\begin{equation}
\frac{dN}{dtdV} \approx \frac{(eE)^2}{16\pi^2} e^{-\frac{\pi m^2}{eE}}.
\end{equation}
\noindent once $E \geq E_{crit} \approx
\frac{1}{\alpha^{1/2}} m{_e}{^2} \approx 10^{18}$ volts/cm.Here we have a
vacuum
breakdown even if the lightest charged particle had a mass as high as
$m_X \approx 10^4$
GeV.

{\bf (2) Linear Accelerators}:  The synchrotron radiation losses for a Planck
energy electron or proton in a circular orbit are prohibitive: even if we take
$R = r_{Hubble} \approx 10^{26}$ meters, we find from \cite{Jac}:
\begin{equation}
\delta W \,\, \mbox{(Synchrotron \,- \, loss)} \approx \left(\frac{E}{m}\right)
^4 \frac{1}{R}
\end{equation}
\noindent that a Planck energy electron (or proton) loses all its energy in
traversing only $10^{-25}$ (or $10^{-14}$) of a turn!
  We should therefore consider (truly straight!) linear accelerators producing
constant gain $G = \frac{\Delta W}{\Delta x}$ with minimal radiation losses.
Since the electromagnetic fields originate in surface charges and currents,
$G_{max} \approx  \frac{e.V.}{A^0} = \frac{10 GeV}{meter} \sim Ry/a_{Bohr}
\approx \frac{1}{2} m{_e}{^2} \alpha^3$ is a maximal gain
allowed.  ($R_y$ = 13.6 ev is the Rydgberg energy constant and $a_{Bohr} =
0.55A^o$ is the Bohr radius).  E fields of such a magnitude destroy materials
by ``skimming" electrons from the top of the Fermi sea.  This implies that
the minimal acceleration length required in order to achieve planckian
energies is  $L_{min} = \frac{m_P}{G_{max}} \approx 10^{20}$ cm
$\sim$30 Parsecs. (Present technology is limited by discharge breakdown at
local defects to $G^{\prime}_{max} \sim \frac{few \; MeV}{meter}$ and the
required $L^{\prime}_{max}$ is 100 K-Parsecs!).  Such a long cylinder is
unstable with respect to bending under tydal forces and/or self gravity.
If bent slightly into a (roughly) circular arc of height $h$, the
prolongation is
$\Delta _L \approx \frac{h^2}{3L}$ (see fig 2) .The corresponding elastic
energy is
\begin{equation}
\Delta E_{Ela} \approx N(Ry) \left(\frac{\Delta L}{L}\right) ^2 \approx
\frac{M}{m_N} \frac{1}{2} m_e \alpha^2 \left(\frac{h^2}{3L^2}\right) ^2
\end{equation}
\noindent with $Ry \approx \frac{1}{2} m_e\alpha ^2$ representing a bond energy
($\left(\frac{\Delta L}{L}\right)^2 Ry$ is the penalty for $\frac{\Delta
L}{L}$ stretch) and $N = \frac{M}{m_N}$ [with $M(m_N)$ the mass of the
cylinder (nucleon) respectively] is the number of bonds.If $\Phi ^{\prime
\prime}_G \approx \frac{1}{r_{Hubble}^2}$ is the local $\vec{g}$ gradient
\footnote{more
appropriately $\phi^{\prime \prime}_G \approx \frac{v{_{escape}}{^2}}{R^2}$
with
$v_{escape}$ the escape velocity from a structure (galaxy, cluster of
galaxies),
etc... of size R.  Since, however, these proper motions such as the infall
towards the Virgo cluster stand up against the Hubble flow
$\frac{v_{escape}^2}{R^2} \approx \frac{1}{r_{H^2}} \approx
\Phi^{\prime \prime}_G$.}
then the tydal energy gain is:
\begin{equation}
\Delta E_{tyde} = M\Phi^{\prime \prime}_G h^2 = \frac{Mh^2}{r_H^2}
\end{equation}
\noindent Also the self gravity gain is:
\begin{equation}
\Delta E_{S.G.} = G_N M^2 \Delta \left(\frac{1}{L}\right)
\approx \frac{M^2}{m{_P}{^2}} \frac{h^2}{L^3}
\end{equation}
\noindent To ensure stability we should keep $\Delta E_{Ela} > \Delta
E_{tyde}$,i.e.  we need
\begin{equation}
h \geq \frac{L^2 \sqrt{m_N/m_e}}{r_H{\alpha}} \geq
10^{16}\, cm
\end{equation}
\noindent where we used $L\geq\L_{min}= 10^{20}$. A weaker bound $h\geq
10^{11}$ cm follows from $\Delta
E_{Ela} \geq \Delta E_{S.G.}$.Even the weaker bound already implies that the
arc in
fig (2-b) is $\theta = \frac{h}{L} \approx 10^{-9}$ rad and a curvature radius
$R
\approx \frac{L}{\theta} \approx 10^{29}\, cm \approx 10r_H$.  From eq (5) and
discussion thereafter, catastrophic synchrotron radiation again follows.
In principle we could attempt to avoid the accelerator pipe, and
have a prearranged set of $N$ acceleration stations $S_1...S_N$ each of
mass $m$, spaced by length $l$ and each accelerating by $\Delta E = m_P/N$.  To
ensure a straight, kinkless, path, we would need to correct each time in the
$(n+1)th$ station the momentum (and location) of the accelerated particle.
Indeed if the ``aperture" of each station is $R$, the emerging particle will
have a transverse momentum uncertainty $\Delta_2 = \frac{\hbar}{R}$ and
the corresponding
angular uncertainty is $\Delta
\theta = \frac{\hbar}{Rm_P}$ .This leads to $\Delta l_{n+1}, \approx \,
Min\, \left\{\frac{\hbar}{m_P}\cdot \frac{l}{R}, R\right\}$ transverse
uncertainty upon arrival at $S_{n+1}$.  To correct for this we may need to
move the magnets, etc. at $S_{n+1}$ by a similar amount. A light signal from
$S_n$ arriving at $S_{n+1}$ a
distance $\delta l = l(1-\beta)  = l\frac{m{_x}{^2}}{m{_P}
{^2}}$ ahead of the accelerated particle could facilitate this providing
$\delta l \geq \Delta l_{n+1}$.  This requires, however, $l \approx
{m_P}^2/{m_X}^2 R \approx (10^{45} - 10^{38})R$ for X =  electron,
nucleon, which is clearly unacceptable.

{\bf (3) Strong Compact and/or Large Scale Magnetic/Electric Fields}

The failure of the linear accelerator to achieve planck energies can be traced
to the fragility of matter which cannot sustain large $E$ fields $E \geq
\alpha^3
m_e^2  \approx Ry/a_{Bohr}$.  Neutron stars are made of much stronger ``nuclear
matter" and ideally could
sustain B fields up to $B_{max} \approx m_N^2 = 10^{19}$ Gauss,
corresponding to energy densities $B^2/8 \pi \approx m_N^4$,.  Due to
the star's rotation (or other
effects), the charged particle accelerated sees an effective electric
field: $E
\approx \frac{v}{c} B \leq B$ over the relevant scale $R$ (size of star, say).
The maximal energy attainable therefore is
\begin{equation}
E_{max} \leq eBR.
\end{equation}
\noindent The total mass of the system $M$  exceeds the magnetic
contribution, $B^2R^3$.  If the system is not a black hole,
then $R_{SW} \equiv \frac{M}{m{_P}{^2}} < R$.  Using $M \geq B^2R^3$, we
find that \begin{equation}
E_{max} \leq em_{Planck}
\end{equation}
\noindent {\it i.e.} any  system producing E.M. fields all the way from
neutron stars to galaxies, or any part of the universe, fails by a factor
$\sqrt{\alpha} \approx \frac{1}{12}$ to obtain Planck energies.

The most energetic cosmic rays observed to date have
energies $\approx 3.10^{11}$ GeV.  These presumably are accelerated on large
cosmological scale by weak magnetic fields, or on short neutron star
scales. In the first case the
maximal energy is limited to $\sim 3.10^{11}$ GeV due to the collision with a
$3^{\circ}$ background photons.

 {\bf (4) Acceleration by Repeated Particle Collisions}

Consider next the cascade of collisions sketched in fig 3.  We start with many,
$N=2^k$, particles all with the same energy $E_0 >m$ and all momenta pointing
roughly to a common point.
Pairs of these particles collide: \#1 with \#2, \#3 with \#4, etc..., \#2$^k$-1
with \#2$^k$.  All collisions are elastic at low center mass energies
and the emerging particles, say \#1 and \#2, etc., have
isotropic angular distributions in the CMS frame of the corresponding pair.
This reflects in a uniform distribution of the energy of emerging particles
$E{_2}{^{\prime}}$, say in the interval $0 \leq E_2^{\prime} \leq
2E{_0}$.
  Let us assume that $E_2^{\prime}\geq E_1^{\prime}, E_4^{\prime} \geq
E_3^{\prime},... \, E{_{2^k}}{^\prime} \geq E^{\prime}_{ 2^k -1}$ (this can
always be achieved by relabeling).  The expectation values of the energies of
each of the more energetic particles in a pair $E_2^{\prime}, E_4^{\prime}$,
etc. averaged over many complete ``experiments" is then $\frac{3}{2}E_0$.  We
will make the (drastically!) simplifying assumption that $E_2^{\prime} \,
E_4^{\prime}$...etc. do in fact always assume their average values, {\it i.e.}
$E_2^{\prime} = E_4^{\prime} =...\, E_{2^k}^{\prime} = \frac{3}{2}E_0$.  We
next
arrange for particle $2^{\prime}$ and $4^{\prime}$, $6^{\prime}$ and
$8^{\prime}$, etc. to collide.  Under the same simplifying assumption we have
for the next generation of particles emerging from this second (k=2) series of
collisions with energies
$E_4^{\prime \prime} = E_4^{\prime \prime}=... \,
E^{\prime \prime}_{2^k} = \left(\frac{3}{2}\right) ^2E_0$.  We will
label $E^{\prime \prime} = E^{[2]},\, E^{\prime \prime
\prime} = E^{[3]}$, etc.
  This process continues for k-1 stages.  At the last (kth) stage particle
$2^{k^{[k-1]}}$ with energy $\left( \frac{3}{2} \right) ^{k-1} E_0$
collides with particle $(2^{k-1})^{[k-1]}$ with the same energy
producing finally $E_f = \left( \frac{3}{2} \right) ^kE_0 \geq m_P$.

Clearly what we attempt here is a highly schematic (and idealized!)
implementation of the abstract concept of ``nested accelerators" of fig 1.
However, as we show next, even this realization fails on physical, kinematical
and
geometrical  grounds.

In order to achieve $\gamma_{final} \approx 10^{22} - 10^{19}$ (for $m =
m_{electron}$ or $m_{nucleon}$ respectively) we need $\left( \frac{3}{2}
\right)^k \geq 10^{20}$.  For each energetic particle with energy $E^{[k+1]}$
we have two colliding ``parent" particles in the (kth) generation.
Furthermore, we require some extra minimal number $\nu$ of ``guiding" or
control ``particles" needed to ensure that indeed the more energetic particles,
say $E_2^{\prime}$ and $E_4^{\prime}$ will collide, rather than $E_2^{\prime}$
and $E_1^{\prime}$, $E_2^{\prime}$ and $E_3^{\prime}$, $E_4^{\prime}$ and
$E_1^{\prime}$, $E_4^{\prime}$ and $E_3^{\prime}$.  (Clearly without this extra
guidance we will have a stochastic thermalized system.)  It would seem highly
conservative to assume $\nu \geq 4$.  Already in this case the total number of
particles $N=(2 + \nu )^k \geq 6^k\geq 10^{20ln_{3/2}^6} \geq 10^{84}$ -
exceeding the total \# of particles in the observed universe.  To avoid
inelastic
collisions, we need to have the momenta of colliding particles very parallel
$\theta ^{[l]} \leq m/E^{[l]} \leq E_0/{(E_0)\left( \frac{3}{2} \right) ^l}
\leq
\left( \frac{3}{2} \right)^{-l}.$

It is very difficult, under these circumstances, to ensure that only the
``chosen" particles will be within interaction range $b_0$ so as to avoid
simultaneous multiple collisions which will ``dilute" the obtained energies by
mixing in low energy particles.Even the demand that initially the $N$
colliding particles are not within each
others' interaction range - a feature always implicitly assumed - is not
trivial.  It implies $R_0 \geq \sqrt{N}b_0$ with $R_0$ the transverse size of
the initial,first generation, beam of particles.  For $b_0 \geq$ fermi the
minimal purely nuclear interaction range, and $N \geq 10^{82}$ the large
minimal required number ($\approx$ total \# of particles) in the universe, we
find $R_0 \geq r_{Hubble}$ a coincidence embodying Dirac's large
number hypothesis.  The transverse focusing (see the expression for
$\theta$ above) implies then
that the length L of the experimental set up,
$L >> R_0 \geq r_{Hubble}.$

\section  {Gravitational acceleration}

Gravitation is, in many ways, the strongest rather than the weakest
interaction. This is amply manifest in the gravitational collapse to a black
hole which no other interaction can stop. Along with the definition of
$ m_P$ this naturally leads us to consider gravitational accelerators, and
the acceleration (or other effects) of black holes in particular.
  If a particle of mass $\mu$ falls to a distance r from a mass m, it
 obtains, in the relativistic case as well, a final velocity
\begin{equation}
\beta_f=\sqrt{r_{SW}\over r}
\end{equation}
  with $r_{SW}={G_N m\over c^2}$, the Schwartzschild radius of the mass m.
In order to obtain in a ``single shot'' planckian
energies, we need that $\epsilon=1-\beta_f=\frac{1}{2\gamma_f^2}=\frac
{\mu^2}{2m_P^2}$ or $\epsilon=\frac{r-r_{SW}}{r_{SW}}
=\frac{\mu^2}{2m_P^2}.$  The last equation applies also if at infinity
we have initially
 a photon of energy$\mu$.  Taking generically
$\mu=GeV=m_N$ and  $ m=m$(neutron star)=$m_{Chandrasehkar}=1.4 m_o=\frac
{m_P^3}{m_N^2}$ as the mass neutron star or black hole doing the acceleration,
we find $r-r_{SW}=l_P=10^{-33}cm $. The distance of closest approach $r$
must exceed $r_0$, the radius of the star (compact object) and hence we have
$r-r_0=l_P$. Thus the system of interest is most likely a black hole. In turn
the black hole will also capture the accelerating particle.

A sophisticated accelerator using repeated sling shot kicks in a system of
four black holes was suggested by Unruh. This beautiful concept is best
illustrated in the following
simple two black holes context. Consider first two black holes of equal mass
$m_1=m_2=m$  at points $P_1 $ and $P_2$ located at $ +L,-L$ along the
$z$ axis.  A relativistic particle $\mu$ (which could also be a photon) is
injected with some relative impact parameter parallel to the z axis near z=0.
With an appropriate choice of the impact parameter $b=b_0$, the accelerated
particle describes a ``semi-circle'' type trajectory around $m_1$ at $P_1$,
and is reflected around this mass by an angle $\theta=\pi$ exactly.
Moving then along a reflected ($ x\to -x$) trajectory the particle $\mu$
approaches the other mass $m_2$ at $P_2$, and is reflected there by
$\theta=\pi$ as well. The particle will eventually describe a closed geodesic
trajectory bound to the two mass $ m_1, m_2$ system.  In reality the two
masses move. For simplicity consider the case when the masses move towards
each other with velocity $\beta$. Transforming from the rest mass of $m_2$
say to the ``Lab frame'', we find that in the reflection the energy of $\mu$ is
enhanced according to $E_\mu \to  E_\mu\sqrt{(1+\beta)\over (1-\beta)}$.  The
last equation represents the boost due to the sling shot kick alluded to above.
If we have $N$ such reflections, the total boost factor
is $\sim(1+\beta)^N$.  This overall boost could exceed $10^{19}$
for $\beta=1/3$ if $N$ exceeds $150$.  However, in this simple geometry the
total number of reflections is limited by $N=1/\beta$.
After more reflections, the two masses will either coalesce or reverse their
velocity, leading now to a deceleration of the particle $\mu$  upon each
reflection. The total amplification of the initial energy is therefore limited
in this case simply to a factor $e\simeq 2.7$.

The Unruh set up involves,
however, two additional heavier black  holes $M_1=M_2=M $with $m_1,m_2$,
revolving around $M_1,M_2$ respectively, in circular orbits of equal radius
$R$ and period $T=2\pi R/\beta$, with $\beta $ the orbital
velocity. The two orbits are assumed to lie in the $ (x-z)$ plane with the
centers of the circles at $(x,z)=(0,+L)$. The
tops of the two circles, {\it i.e.} the points where x is maximal, define now
the original reflection centers  $ P_1,   \ \ P_2 = (R,-L),\ \  (R,+L) $.
The oppositely rotating masses $m_1$ and $m_2 $ are synchronized to pass at
$P_1$ and $P_2$, respectively, at the same time -- once during each period T.
Furthermore, the motion of the accelerated mass $\mu$
is timed so as to have $\mu$ at the extreme left point on its ``Stadium
Shaped'' orbit $(x,z)=(0,-L-\rho)$ or, at the extreme right point
$(x,z)=(0,L+\rho)$, at precisely the above times.  This then allows us to
achieve the desired sling shot boosts, repeating once every period T. Note that
2T is now the period of the motion of the five body system $(M_1,M_2,m_1,m_2
;\mu)$.

However, the inherent instability of this motion again limits the
number $N$ of periods (and of sling shot boosts)
and,as pointed to us by B. Reznik,  foils this ingenious device.
The assumed hierarchical set-up $ L>> R>>r_{SW}=b_0$ can be used to approximate
the angular deflection of $\mu$ while it is circulating around $ m_1 $, say by
\begin{equation}
\theta = \int_{u_{min}\simeq0}^{u_{max}} {du \over
\sqrt{ {1\over b^2} - u^2( 1-2G_N m u) }}
\end{equation}
with $b$ the impact parameter and $u_{max}=1/\rho$ corresponding to the
turning point of closest approach. Independently of the exact (inverse eliptic
function) dependence of $ \theta$ on $b/{r_{SW}},\ \rho/{r_{SW}}$, we expect
that a fluctuation $\delta b^0$ around the optimal $ b^0$, for which $\theta$
equals $\pi$, causes a corresponding fluctuation in $\theta$ : $ \delta^1\theta
=\pi-\theta=c\delta b^0/b^0$, with the dimensionless constant $c$ being of
order one. The large distance $L$ transforms this small $\delta\theta$ into
a new impact parameter deviation, $\delta^1(b) = L\delta^1\theta$. The ratio
between successive impact parameter is then given by $|\delta^{1}(b)|=(cL/b_0)
\delta^0(b)$. After $N$ reflections, we have therefore
\begin{equation}
\delta^N(b) \simeq (L/b_0)^N \delta^{(0)}(b)
\end{equation}

For $ L>R>b_0$ we expect a large growth rate of the fluctuations in the impact
parameter.  The circular trajectory of m, with radius R, decays at a rate
proportional to $\beta ^5$ due to gravitational radiation. This limits
$ \beta$ and implies that we need a large number (of order 100) reflections
to achieve $10^{19}$ energy enhancement.  This implies that in order to avoid
complete orbit deterioration for the accelerating particle $\mu$, {\it i.e.}
to avoid $\delta^N(b)\simeq b_0$, we need unattainable initial precision
$\delta b^0/{b_0}=10^{-100}$, which in particular exceeds the quantum
uncertainty .

Ultra-relativistic acceleration can be achieved in the universe as a whole; the
Hubble velocities of the most distant galaxies or oldest particles,
$\beta_{H}$, are arbitrarily close to one in a sufficiently large and flat
universe, though no collisions at planck CM energies can be thus engineered.
  It is amusing to recall one ``simple" context in which
Planck energies may be achieved but not much exceeded -- namely the evaporating
mini-black holes.  Indeed, as indicated in the Introduction, black holes emit
all elementary quanta with a thermal spectrum.  Only as $R_{BH} \arr l_P$ does
$T_{BH} \arr m_P$, so that a Planckian black hole would have generated Planck
energy quanta except for the fact that at this point $M_{BH} \arr m_P$ as well,
and energy conservation prevents achieving super-planckian energies.

 If as it emits the last quanta, the center of
mass of the black hole had appreciable boost, say $\gamma \geq 3$, then one
might expect that Hawking quanta emitted in this direction would have
super-Planckian energies: $\gamma E_{\mbox{in rest frame}}\geq m_P$.
The recoil momentum accumulated through
the Hawking radiation is, at \underline{all} stages: $P_{Rec}
\approx m_P$, so that $\gamma_{Rec} \sim 1$, and
no appreciable extra boost effect is expected.  $P_{Rec} \approx m_P$ implies
a recoil kinetic energy of the black holes $F_{recoil} \approx \frac{P^2}
{2M_{BH}} = \frac{M{_P}{^2}}{2M_{BH}} = T_{BH}$ as required by equipartition.
{This fact features in a new approach to Hawking radiation which
attempts to avoid any reference to transplanckian acceleration due to
B. Reznik}.

\section{Summary and Further Speculations}

The difficulty of conceiving even Gedanken super-planckian
accelerators
suggests that models without superplanckian elementary particles may be
consistent.  In particular models with light fermions may not  allow
acceleration to super-planckian energies
when embedded into the present universe with its given Hubble radius.
This could not happen in an infinite or sufficiently large open
universe.However the  universe with its
specific global parameters does provide the required violation at the
present time. Our discussion was clearly not exhaustive and the
possibility that some ingenious suggestion (which we failed to
realize), can actually lead to a planck accelerator is still open.  A
more comprehensive discussion is presently under preparation and will
address some further possibilities.
Let us conclude with several comments and speculations.

a) Our considerations do not exclude theories with super-planckian energies
and masses of elementary excitation.Indeed the very notion of what is an
elementary particle may be profoundly revised in schemes such as string
theory.  Rather, all that is hinted is a possible
consistency of theories where such a super-planckian regime is unphysical and
is excluded.

(b) It has been suggested that ``Planck Scale Physics" induces
effective interactions violating all global symmetries.  A particular
example is a $\frac{\lambda}{m_P} \Phi^+ \Phi \Phi^+ \Phi \Phi$ term where the
$\Phi$ bosons carry two units of lepton number.  Such a term violates $U(1),
\ \ (B-L)$ and therefore endows the corresponding would-be massless Goldstone
boson (Majoron) with a finite mass.  A concrete mechanism involves the
formation of a black hole in a collision of, say, $\Phi^+\Phi$, followed by
the decay of the B.H. into $\Phi \Phi \Phi^+$, a final state with two units of
lepton number. In this way the violation of the global quantum numbers traces
back to the fundamental ``No Hair Theorem" for black holes.  Exactly as in
the case of SU(5), where a virtual $X,Y$ GUTS meson can mediate nucleon decay
by
generating effective four Fermi terms, the virtual ``mini black hole"
system was conjectured to induce the $\frac{\lambda}{m_P} \Phi^+ \Phi \Phi^+
\Phi \Phi $ term.  The estimated resulting Majoron mass $M_x \approx KeV$ is
rather high.\cite{Moha}
Also Planckian black holes would constitute some irreducible environment
and may require modification of quantum mechanics\cite{Rez}
If, , the whole super-planckian domain is inaccessible for
elementary excitations, it is conceivable that  such effects may not
be there.  Super-planck physics --
even in terms of indirect low energy manifestation would then be
completely absent.

(c) `tHooft has been emphasizing \cite{`tHooft} that understanding the
elementary
particle--black hole connection issue may lead to a more profound understanding
of both field theory and gravity.
Our approach attempts to  realize the 'tHooft conjecture, but in a very
different way.  Rather than try to bridge the gap between particles and black
holes, we elevate such a gap -- or the upper bound $E \leq m_P$ on the
energy-momentum of any elementary excitation -- to a fundamental postulate of
the theory.
The basic $\Delta x \Delta p \geq \bar{h}$ postulates, allows a qualitative
understanding of the spectra and structure of atoms.  Our hope is that $E,\, P
\leq  m_P$,  will allow better understanding of elementary particles,
cosmology, and their interrelations.

(d) As indicated above
 collisions of elementary particles at such energies do occur, but
very rarely, with exponentially small probabilities.  However, when such a
collision occurs, there could be potentially  dramatic repercussions.
The newly formed mini
black hole could be a region of a new expanding ``baby universe" where the very
basic fundamental parameters may be different.  Indeed, we  would like
to suggest that {\it a
posteriori} explain the rarity of such collisions, which spell the ``end" of
the
present universe.  The new universe still preserves the same gauge group $SU(3)
\times SU(2) \times U(1)$ or a corresponding $SU(5)$ or $SU(10)$, etc. GUTS
group.  The point is that these small length scales( $R_{BH} \approx
l_P$) all symmetries are restored, and the ``gauge hair" is a common link
between the two universes.  New fermion masses  and even new fermionic
degrees of freedom may arise in this
process, but due to the ``gauge memory'' it is natural to assume the same gauge
structure for them.  This could then provide an ``evolutionary" explanation for
the repeating fermionic generations in the spirit of  speculation
by Nambu\cite{Namb} and Coleman\cite{Cole}.

Indeed, as indicated by our estimates the new generations with lighter
electrons
and quarks could make it more difficult to achieve super-planckian energies
again in the new universe, thus allowing  the long lived present
universe. If fermion masses stay fixed and the universe expands for
ever(which appears to be observationally more favorable at the
present time)then planck accelerators would become eventually
possible.  The above extremely heuristic notion, could
potentially also evade this difficulty.    A  continuous change of
fundamental constants with the
expansion of the universe  was suggested by the Dirac large number
hypothesis. This clearly failed various experimental checks.  Our conjecture
that  a universe which is  one billion times
older than the present universe would still fail to accelerate to
planck energies due to the possible emergence of a fourth superlight
generation is much harder to check.

{\bf Acknowledgments}

 We are indebted to P. Mazur for patiently explaining to us his novel
approach. The encouragement and useful comments of Y. Aharonov, A.
Falk, O. W. Greenberg,
J. Knight, S. Migdal, H. Neuberger, S. Polyakov, L. Susskind, P.
Steinhardt, and E. Witten, are gratefully acknowledged.  We are
particularly indebted to W. Unruh who suggested the
 ``Four Black Hole'' Planckian Accelerator  constituting
the most sophisticated suggestion to date, and (even more so) to B.
Reznick who found the   instability   which foiled the Unruh accelerator.

\begin{center}{\large \bf Figure Captions}
\end{center}

\begin{enumerate}
\item The nested ``Accelerator Within Accelerator" system designed to achieve
super-planck energies.

\item A Gedanken super-planck linear accelerator of length {\it L} and radius
{\it R}.  It is bent into a circular arc of angle $\Delta \theta$ (not
indicated).  The center of the chord to the arc
is at a height {\it h} below the middle of the arc.

\item A multiparticle collider designed to ``Breed" -- by repeated collisions
and choices of the more energetic particles to collide in consecutive stages --
a super-planck particle.

\item The Unruh accelerator.  The two black holes $m_1 = m_2 = m$ go around say
the stationary more massive $M_1 = M_2 = M$ black hole in circular orbits of
radius {\it R} and in opposite direction.  The accelerating particle goes
around in the oblong ``stadium-like" trajectory of thickness $2b_o$ with $b_o$
the impact parameter.  It gets the ``sling-shot kicks" boosting its energy as
it goes around $P_1,\, P_2$ at times $t, t + T$ with $m_1,\, m_2$ at $P_1,\,
 P_2$ respectively.

\end{enumerate}

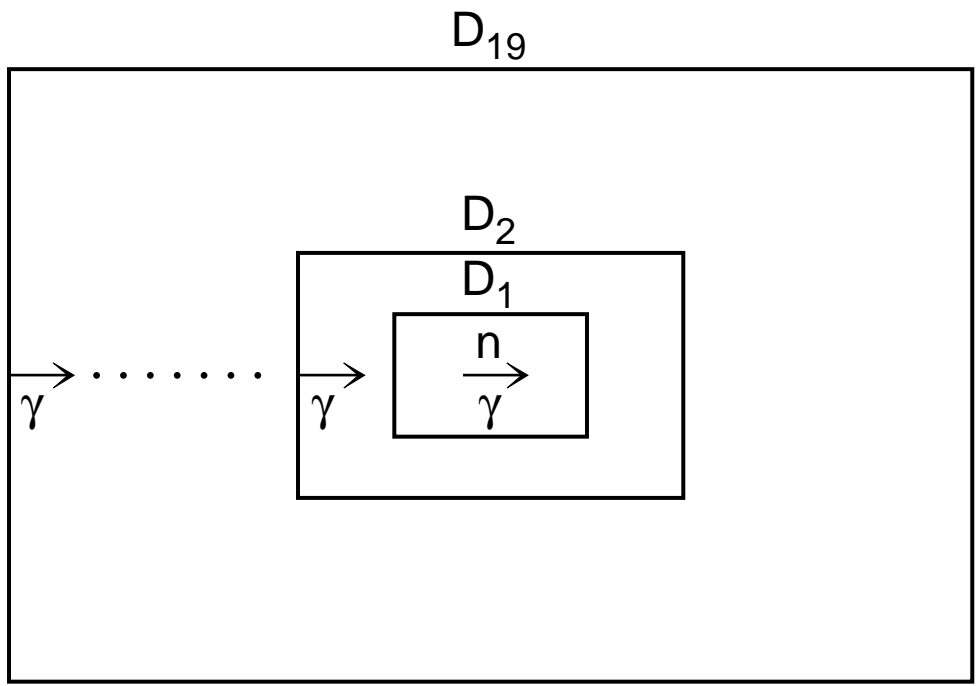
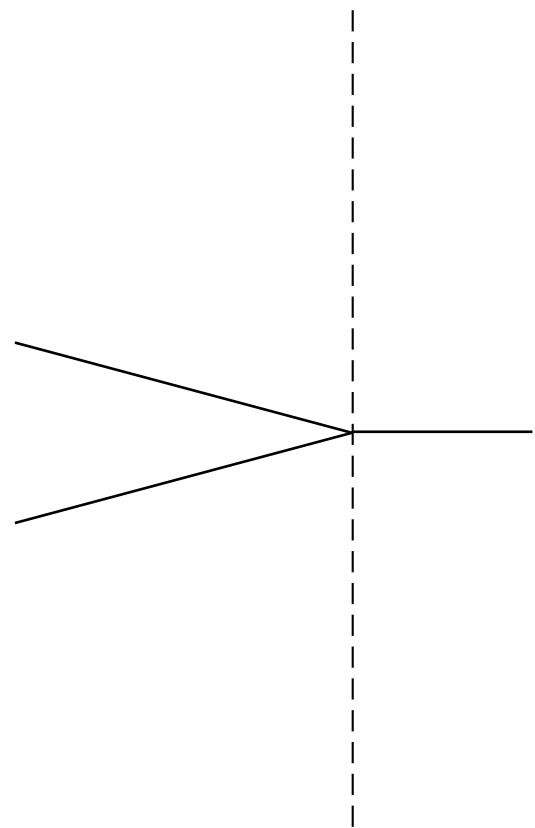
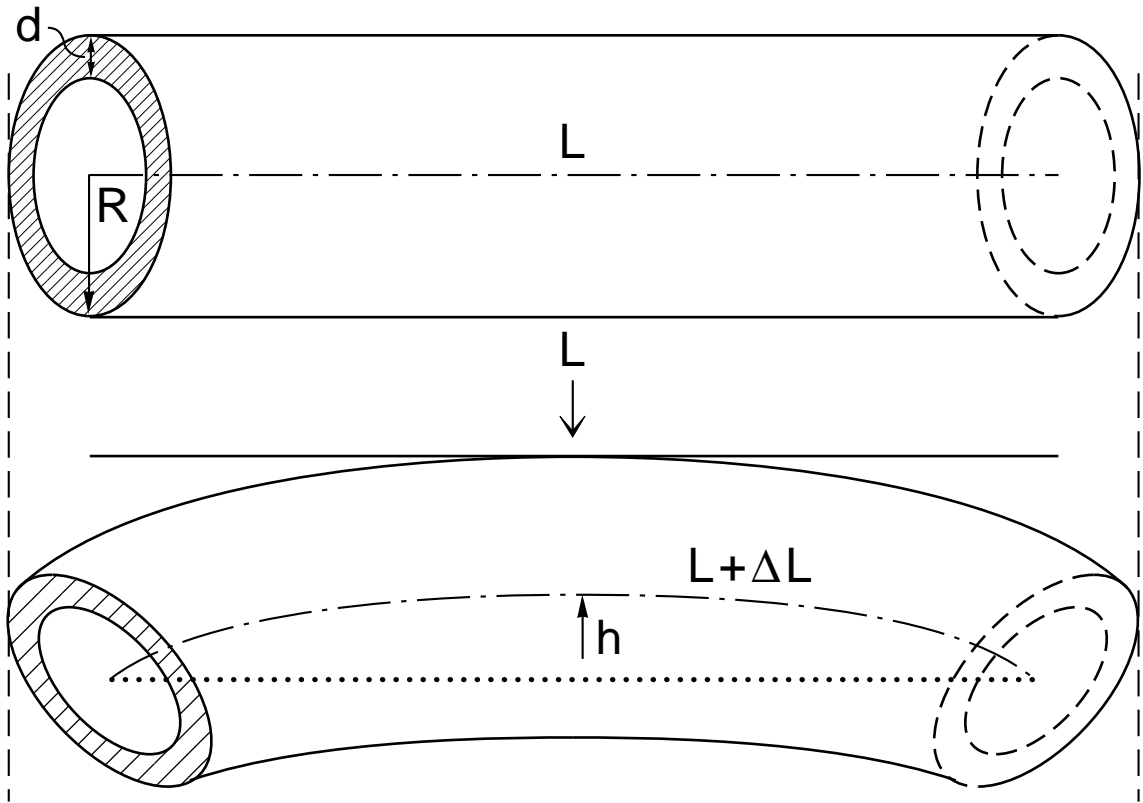
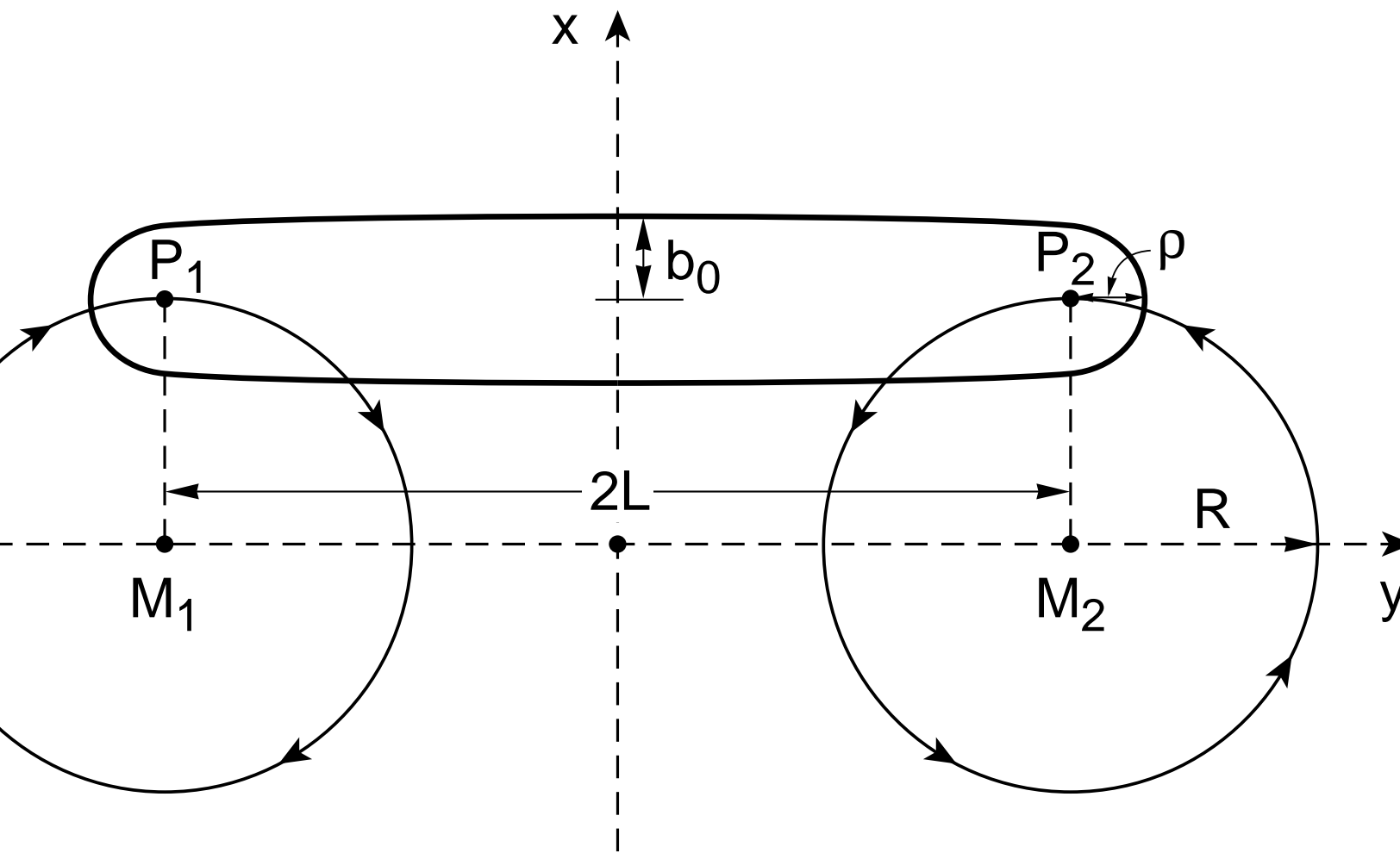
\end{document}